\documentclass[preprint,showpacs,preprintnumbers,amsmath,amssymb,nofootinbib]{revtex4}

\usepackage{graphicx}
\usepackage{dcolumn}
\usepackage{bm}

\newcommand{\bi}{\bibitem}

\newcommand{\be}{\begin{eqnarray}}
\newcommand{\ee}{\end{eqnarray}}
\catcode`\@=11
\def\lsim{\mathrel{\mathpalette\@versim<}}
\def\gsim{\mathrel{\mathpalette\@versim>}}
\def\@versim#1#2{\vcenter{\offinterlineskip
\ialign{$\m@th#1\hfil##\hfil$\crcr#2\crcr\sim\crcr } }}
\catcode`\@=12

\begin{document}

\preprint{KANAZAWA-06-15}

\title{The SUSY Flavor Problem, Proton Decay \\ and \\
Discrete Family Symmetry
\footnote{Talk given by J.K. at SUSY06, Irvine, California, 12-17 June 2006.}
\vspace{1cm}
}

\author{Etsuko Itou$^a$, Yuji Kajiyama$^b$ and Jisuke Kubo$^c$
\vspace{1cm}}

\affiliation{$^a$ Department of Physics, Graduate School of Science,
  Osaka University, Osaka 560-0043, Japan\\
  $^b$ National Institute of Chemical Physics and Biophysics,
Tallinn 10143, Estonia\\
 $^c$ Institute for Theoretical Physics, Kanazawa University
Kanazawa   920-1192, Japan
\vspace{3cm}}

\begin{abstract}
We consider a supersymmetric
extension of the standard model, which possess a family
symmetry based on a binary dihedral group
$Q_6$, and investigate the consequences of the
family symmetry on the  mixing
of fermions, FCNCs and the stability of
 proton.
 \vspace{1cm}
\end{abstract}


\maketitle

The classification of finite groups has been completed 
1981 by Gorenstein,
about 100 years later than the case of the continues group.
Therefore, we believe it is worthwhile to look at finite groups
more in detail and find applications into particle physics.
In fact there are renewed interests  \cite{babu0,ivo}  in finite groups 
as such as $S_3$ or $A_4$  to explain the large mixing of
neutrinos.

In \cite{babu1} it has been motivated to obtain
a mass matrix of the nearest neighbor type \cite{weinberg}
from a non-abelian discrete family symmetry.
We found that this is in fact possible and that the smallest
group is the binary dihedral group $Q_6$,
which is the covering group of the smallest non-abelian
group $S_3$.
There are two two-dimensional irreps of $Q_6$;
${\bf 2}_1$ is pseudo-real 
 and ${\bf 2}_2$is a real irrep.
There are also two real one-dimensional irreps.
${\bf 1}_{+,0}, {\bf 1}_{+,2}$, 
and two
complex one-dimensional irreps. ${\bf 1}_{-,1}, {\bf 1}_{-,3}$,
while ${\bf 1}_{+,0}$ is 
the true singlet \cite{frampton1}.
Table I shows the $Q_6$ assignment.
This is an alternative assignment to the one given in
 \cite{babu1}. 
We consider  this assignment, because
it can explain the  maximal mixing 
of the atmospheric neutrinos.
(The leptonic sector is basically the same as
that of \cite{kubo4}.)
Further, we assume that CP is spontaneously broken.
It is possible to construct a Higgs superpotential
for which CP can be spontaneously broken.
\begin{table}[thb]
\begin{tabular}{|c|c|c|c|c|c|c|c|c|c|c|c|}
\hline
 & $Q$ 
 & $Q_3$  
& $U^c,D^c$  
& $U^c_3,D^c_3$ 
 & $L$ & $L_3$ 
 &$E^c,N^c$ & $E_3^c$  &   $N_3^c$ 
  & $H^u,H^d$
 & $H^u_3,H^d_3$ 
\\ \hline
$Q_6$ &${\bf 2}_1$ & ${\bf 1}_{+,2}$ &
 ${\bf 2}_{2}$ &${\bf 1}_{-,1}$ &${\bf 2}_{2}$ &
${\bf 1}_{+,0}$  & ${\bf 2}_{2}$ & 
${\bf 1}_{+,0}$ & ${\bf 1}_{-,3}$ &
${\bf 2}_{2}$ & ${\bf 1}_{-,1}$   \\ \hline
\end{tabular}
\caption{$Q_6$ assignment.}

\label{assignment}
\end{table}

\begin{figure}[htb]
\includegraphics*[width=0.4\textwidth]{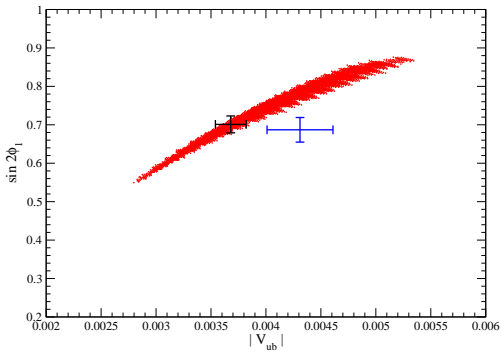}
\caption{\label{fig1}\footnotesize
Predicted area in the $|V_{ub}|-\sin2\phi_1(\beta)$ plane.
The results of UTfit group \cite{utfit} (black)
and of Particle Data Group 2006 \cite{pdg} (blue)
are plotted, too.
}
\end{figure}

In the quark sector we have 9 independent real parameters
to describe 6 quark masses and 4 parameters of the
CKM matrix. So there is one real prediction,
which can be displayed in different planes.
The absolute value of $V_{td}$ over
$V_{ts}$, for instance, is predicted
to be $0.23\pm 0.02$ by the model.
This  can  be directly compared with
the experimental values $0.16\pm 0.04$ \cite{pdg}
and $0.208\pm 0.07$ \cite{utfit}
because the oscillation frequency in the
$B_s-\bar{B}_s$ system has been  measured at Tevatron
this year \cite{tevatron}.
Fig. \ref{fig1}
shows the prediction of the model in the $V_{ub}-sin 2\phi_1(\beta)$
plane. 

In the leptonic sector, there are only
7 independent real parameters, of which one is a CP phase,
to describe 6 masses of the charged leptons and 
neutrinos, three angles of the
neutrino mixing matrix along with
one Dirac CP phase and three Majorana phases.
So there are 12-6=6 predictions.
First: the model predicts the neutrino mass spectrum
is inverted.
Second: there exist only one independent CP phase, and 
the absolute scale of the neutrino mass 
depends on  the independent phase.
Third:
$|U_{e3}|=m_e/\sqrt{2}m_\mu+O(10^{-5}) 
\approx 0.0034$,
and $|U_{\mu 3}|=1/\sqrt{2}+O(10^{-5})$.
Fourth:
Since there exits only one independent phase,
the average neutrino mass appearing in neutrinoless
double beta decays can be predicted as a function of 
the independent phase.
For a wide range of the independent phase,
the average mass stays at the minimum$
( 0.034  - 0.069 ) ~~\mbox{eV}$ \cite{kubo4}.

Now we come to proton decay \cite{itou}.
As we know, the lowest dimension
of the proton-decay-leading
operators is five, if we assume 
$R$ parity \cite{protondecay}.
With R parity, there are two types
of operators, the left-handed and right-handed types \cite{protondecay}.
If there are no further constraints,
there will be 27 independent operators for each.
These operators can be generated by GUTs
or by some unknown Planck scale physics.
Here we assume that  unknown Planck scale
physics generates baryon number violating
operators and respects the $Q_6$ family symmetry.
Then it turns out that the number of the independent
left-handed operators reduces to ONE and in the case of the
right-handed operators to TWO. 
Moreover, the left-handed operator 
\be
\sum_{I=1,2}Q_I Q_I Q_3 L_3
\label{op}
\ee
(see Table 1) gives
the dominant contribution, so that in the first approximation
there is only one coupling constant. 
The reason that the left-handed operator gives the dominant
contribution to proton decay
is basically the same as the in the usual case 
\cite{belyaev} without
family symmetry.
The usual argument \cite{belyaev} is based on the fact that 
the diagrams with the neutral gaugino loops
are negligibly suppressed if all the 
squarks masses are degenerate.
The degeneracy of the squark masses is
needed to suppress FCNCs.
This cancellation mechanism does not work for
the diagrams with  wino loops.
Therefore, the left-handed operators,
which can be dressed with the  wino loops,
are dominant over the right-handed operators.
As we will see later on, the soft scalar mass matrices are diagonal,
 and the first and second elements are equal by 
 the family symmetry.
So, the almost degeneracy of the squaks masses
is automatic thanks to the family symmetry.

Here we would like to explain some structure of the left-handed 
operator (\ref{op}).
The quark superfield $Q_3$ which is singlet of $Q_6$ contains
only little component of the u and d quarks,
which can be read off from $|U_{uL}|\simeq 0.0023$,
where $U_{uL}$ is the mixing matrix of the left-handed quarks.
This gives an overall suppression of
$10^{-6}$
in the decay mode into a charged lepton.
The rate of 
 the decay mode into a charged lepton  is controlled by
$(U_{eL})_{\tau 1}=1$ and  $(U_{eL})_{\tau 2}=m_e/m_\mu$, 
where $U_{eL}$ is the mixing matrix of the left handed leptons.
From this observation we conclude that
the branching fraction  for the decay into a muon
is five orders of magnitude smaller than the decay  into a positron.
Moreover, this small number is nothing
but  $|U_{e3}$.

Since there is basically only one coupling constant
for the left-handed operator (\ref{op}) 
in our model, the relative branching fractions
can be fixed in the first approximation,
and can be compared with the result of
the minimal $SU(5)$ 
supergravity model \cite{yanagida,murayama}.
In the SU(5) case,
 it is assumed that the up type quarks
are mass eigenstates from the beginning.
(If this type of assumptions are removed,
one can find different conclusions \cite{perez}.)
We find that  there are considerable 
differences \cite{yanagida,murayama}.

We believe that the effect of supersymmetry breaking appears
as soft-supersymmetry breaking terms in our 4D Lagrangian.
Moreover,  one has to
highly fine tune these parameters so that they do not cause 
problems with experimental observations
on the FCNC processes and CP-violation phenomena.
There are several approaches to overcome this problem.
Here we consider
a mechanism which is based on 
the family symmetry $Q_6$.
In the present model, the soft scalar mass matrices
are diagonal by the family symmetry, and  
the first two entries are the same 
by the same symmetry.
Further, the left-right soft mass matrices have the same structure
as the fermion mass matrices.
(This structure is the same as the $S_3$ invariant
supersymmetric model of \cite{kobayashi}.)
Since we assume that CP is spontaneously broken,  these
soft parameters are real.
We found that  the family symmetry and 
the assumption of the spontaneous CP violation
interplay in such a way that  the CP phases of the 
$\delta$'s cancel exactly,
where $\delta$'s \cite{fcnc-mueg} are dimensionless parameters measuring
the deviation of the corresponding
soft parameters from  the universal ones.
In this way we can satisfy the most stringent constraints
coming from the EDMs.
We have 
calculated the deltas $\delta$'s
and compared with the experimental
 bounds given in \cite{fcnc}.
\begin{table}[thb]
\begin{tabular}{|c||c|c|}
 \hline
 &  Exp. bound  & $Q_6$ Model \\
   \hline  \hline

$|(\delta^e_{12})_{LL}|$
& $4.0 \times 10^{-5} ~\tilde{m}_{\tilde{\ell}}^2$
& $4.9 \times 10^{-3} \Delta a_L^{\ell} $
\\ \hline
 $|(\delta^e_{12})_{RR}|$
& $9 \times 10^{-4} ~\tilde{m}_{\tilde{\ell}}^2$
&$8.4 \times 10^{-8} \Delta a_R^{e} $
\\ \hline
$|(\delta^e_{12})_{LR}|$
 & $8.4 \times 10^{-7} ~\tilde{m}_{\tilde{\ell}}^2$
&$\sim 5 \times 10^{-6}\tilde{m}_{\tilde{\ell}}^{-1}$
\\ \hline
 $|(\delta^e_{13})_{LL}|$
 & $2 \times 10^{-2} ~\tilde{m}_{\tilde{\ell}}^2$
&$1.7 \times 10^{-5}\Delta a_L^{\ell} $
\\ \hline
$|(\delta^e_{13})_{RR}|$
&  $3 \times 10^{-1} ~\tilde{m}_{\tilde{\ell}}^2$
&$5.9 \times 10^{-2}\Delta a_R^{e} $
\\ \hline
 $|(\delta^e_{13})_{LR}|$
& $1.7 \times 10^{-2} ~\tilde{m}_{\tilde{\ell}}^2$
&$\sim 3 \times 10^{-7}\tilde{m}_{\tilde{\ell}}^{-1}$
\\ \hline
\end{tabular}
\caption{Experimental bounds on  $\delta$'s and the theoretical values in $Q_6$ model, where
the parameter
$\tilde{m}_{\tilde{\ell}}$ denote
$m_{\tilde{\ell}} /100$ GeV. 
See \cite{itou} for the quark sector.}
\end{table}
Table 2  shows some examples of the case of the leptonic sector.
The capital deltas $\Delta$'s are free dimensionless parameters
which can not be constrained by the family symmetry.
The small numbers appearing in the right column
have approximate analytic expressions.
For instance, $4.9\times 10^{-3}$  is 
$m_e/m_\mu$, which is  $\sqrt{2}|U_{e3}|$.
In the quark sector,
apart from some cases, the soft supersymmetry breaking
parameters need not be fine tuned in the present model
to satisfy the experimental constraints coming from FCNCs
and CP violations (see \cite{itou} for more details).

To conclude, we could say that 
the smallness of the three apparently independent
quantities has the same origin in the model;
the smallness of $|U_{e3}|$,
the suppression of $\mu\to e\gamma$,
and the ratio of  proton decay branching
fractions into a $\mu$ and a $e^+$.
This is a consequence of a low energy flavor symmetry.


\begin{thebibliography}{99}
\bibitem{babu0}
K.~S. Babu, E.~Ma and J.W.F.~Valle,
 Phys. Lett. {\bf B552,} 207 (2003);
G.~Altarelli and  F.~Feruglio,
Nucl. Phys. {\bf B720,} 64 (2005);
A~.Mondragon,
hep-ph/0609243.
\bibitem{ivo}
I.~de Medeiros Varzielas,
Talk given at this meeting,
hep-ph/0610351, and
I. de Medeiros Varzielas, S.~F. King
and G.~G. Ross, hep-ph/0607045.

\bibitem{babu1}
K.S.~Babu and J.~Kubo, Phys. Rev. {\bf D71,} 056006 (2005).

\bibitem{weinberg}S.~Weinberg,
in {\em Transactions of the New York Academy of Sciences}, 
Series II, Vol. 38, 185 (1977);
 F.~Wilczek and A.~Zee,
Phys. Rev. Lett. {\bf 42,} 421 (1979);
H.~Fritzsch,
Phys. Lett. {\bf B 73,} 317 (1978);  Nucl. Phys. {\bf B155,} 189 (1979). 
\bibitem{frampton1}
 P.H.~Frampton and T.W.~Kephart,
Int. J. Mod. Phys. {\bf A10,} 4689 (1995);
Phys. Rev. {\bf D64,} 086007 (2001). 

\bibitem{kubo4}
J.~Kubo, A.~Mondrag\'on, M.~Mondrag\'on and
E.~Rodr\' iguez-J\' auregui,  Prog. Theor. Phys.
{\bf 109,} 795 (2003); Erratum-ibid. {\bf 114,} 287 (2005);
J.~Kubo, Phys. Lett. {\bf B578,} 156 (2004);
Erratum-ibid. {\bf B619,} 387 (2005).

\bibitem{utfit}
M.~Bona {\em et al}., [UTfit Collaboration], hep-ph/0606167.

\bibitem{pdg}
A.~Ceccucci, Z.~Ligeti and 
Y.~Sakai, [Particle Data Group], J. Phys. {\bf G33,} 1 (2006).

\bibitem{tevatron}
D0 Collaboration, Phys. Rev. Lett. {\bf 97,} 021802 (2006);
CDF Collaboration,
 Phys. Rev. Lett. {\bf 97,} 062003 (2006).

\bibitem{itou}
E.~Itou, Y.~Kajiyama and J.~Kubo,
Nucl. Phys. {\bf B743,} 74 (2006),
and references therein.


 \bi{protondecay}
 N.~Sakai and T.~Yanagida,
Nucl. Phys. {\bf B197,} 533 (1982),
S.~Weinberg,
Phys.~Rev.~{\bf D26,} 287 (1982).

\bi{belyaev}
V.M.~Belyaev and M.I.~Vysotsky,
Phys.~Lett.~{\bf B127,} 215 (1983);
J.~Ellis,~S.~Hagelin,~D.V.~Nanopoulos and K.~Tamvakis,
Phys.~Lett.~{\bf B124,} 484 (1983).

\bi{yanagida}
J.~Hisano,~H.~Murayama and T.~Yanagida,
Nucl.~Phys.~{\bf B402,} 46 (1993).
\bibitem{murayama}    
H.~Murayama and D.B.~Kaplan,  
Phys.~Lett.~{\bf B336,} 221 (1994).  


\bibitem{perez}
Talk given by P.~F.~Perez at this meeting, and
B.~Bajc, P.~F.~Perez and G.~Senjanovi\'c, 
Phys. Rev. {\bf D66}, 075005 (2002).

\bibitem{kobayashi}
T.~Kobayashi, J.~Kubo and H.~Terao, 
Phys. Lett. {\bf B568}, 83 (2003),
and references in \cite{itou}.

\bibitem{fcnc-mueg}
L.~Hall, V.A.~Kostelecky and S.~Raby,
Nucl.~Phys. {\bf B267,} 415 (1986);
F.~Gabbiani and  A.~Masiero,
Phys.~Lett. {\bf B209,}  289 (1988).

\bibitem{fcnc} 
F.~Gabbiani, E.~Gabrielli, A.~Masiero and L.~Silvestrini,
Nucl.~Phys. {\bf B477,}  321 (1996);
S.~Abel, S.~Khalil and O.~Lebedev,
Nucl.~Phys. {\bf B606,}  151 (2001); M.~Endo, M.~Kakizaki and 
M.~Yamaguchi, Phys.~Lett.~{\bf 583,} 186 (2003); 
J.~Hisano, Nucl. Phys. Proc. Suppl. {\bf 137,}
169 (2004);
J.~Hisano and  Y.~Shimizu, 
Phys. Rev. {\bf D70,} 093001 (2004).


\end{thebibliography}
\end{document}